\documentclass{PoS}

\usepackage{amsmath,amssymb,amsfonts} \usepackage[english]{babel}
\usepackage{cite}

\usepackage{ulem}

 \newcommand{\A}{\textrm{a}\,}
 
\newcommand{\be}{\begin{equation}} \newcommand{\ee}{\end{equation}}
\newcommand{\bea}{\begin{eqnarray}} 
\newcommand{\eea}{\end{eqnarray}} 
\newcommand{\bmp}{\noindent\begin{minipage}{16cm}}

\definecolor{title}{rgb}{0.20,0.20,0.70}
\definecolor{emph}{rgb}{0.18,0.18,0.60}
\definecolor{emph1}{rgb}{0.18,0.18,0.60}
\definecolor{emph2}{rgb}{0.70,0.18,0.18}

\title{Gradient flow and IR fixed point in SU(2) with Nf=8 flavors}

\ShortTitle{Gradient flow and IR fixed point in SU(2) with Nf=8 flavors}

\author{\speaker{Viljami Leino}\\ Helsinki Institute of Physics and
Department of Physics, University of Helsinki\\ E-mail:
\email{viljami.leino@helsinki.fi}}

\author{Tuomas Karavirta\\ {CP}$^{ \bf 3}${-Origins} \& DIAS, University
of Southern Denmark,\\ 
        E-mail: \email{karavirta@cp3-origins.net}}

\author{Jarno Rantaharju\\
       CP$^{ \bf 3}${-Origins}, IFK \& IMADA, University of Southern
       Denmark\\ E-mail: \email{rantaharju@cp3.dias.sdu.dk}}

\author{Teemu Rantalaiho, Kari Rummukainen, Joni M. Suorsa, Kimmo Tuominen
\\ Helsinki Institute of Physics and Department
of Physics, University of Helsinki\\ E-mail:
\email{teemu.rantalaiho@helsinki.fi}, \email{kari.rummukainen@helsinki.fi}, 
\email{joni.suorsa@helsinki.fi}, \email{kimmo.i.tuominen@helsinki.fi}}

\abstract{We study the running of the coupling in SU(2) gauge theory
with 8 massless fundamental representation fermion flavours, 
using the gradient flow method with Schrödinger functional boundary conditions.
Gradient flow allows us to measure robust continuum limit 
for the step scaling function. 
The results show a clear indication of 
infrared fixed point consistent with perturbation theory.
}
\FullConference{The 33rd International Symposium on Lattice Field
Theory\\ 14 -18 July 2015\\ Kobe International Conference Center, Kobe,
Japan*}

\begin{document}

\section{Introduction} 
There exists a class of asymptotically free gauge theories 
where the running coupling approaches a non-trivial infrared fixed point (IRFP) 
and long-distance physics is conformal.
These theories have applications in model building beyond the standard model, 
particularly in technicolor theories, 
where the electroweak symmetry is broken 
by the dynamical formation of the chiral condensate analogously to QCD. 
In addition to direct applications in particle phenomenology, 
the theories with IRFP are interesting from purely theoretical
viewpoint of understanding the structure of gauge field theories.
In particular, as a function of number of colours $N$, flavours $N_f$ and 
fermion representations, one is interested in determining the boundary 
where the the IRFP of the theory is lost and the behaviour becomes generically QCD-like. 
Where this happens signifies the location of the edge of 
the so called conformal window in the gauge theory phase diagram.
Typically the fixed point coupling near this boundary 
is expected to be large and non-perturbative methods are needed.


In this work we study the running of 
the coupling in SU(2) gauge field theory with 8 fermions 
in the fundamental representation.
Analytic estimates suggest that this theory can be expected to lie within the
conformal window, i.e has an IRFP.
It has been studied before in~\cite{Ohki:2010sr}, 
but the results were inconclusive. The
phase structure of the model 
with staggered fermions was studied in~\cite{Huang:2014xwa}.

The SU(2) gauge theory with 4--10
fundamental representation fermion flavours 
has been studied also earlier~\cite{Bursa:2010xn,Karavirta:2011zg,Hayakawa:2013maa,Appelquist:2013pqa}. 
In these works, 
the coupling was measured via the Schrödinger functional method~\cite{Luscher:1991wu}, 
but the results were inconclusive regarding the location of 
the edge of the conformal window.  
Since the running is slow close to the IRFP, 
very high accuracy is needed to discern the continuum behaviour, 
hindering the progress towards large enough volumes.

In this work we use the gradient flow method~\cite{Luscher:2011bx} 
with Schrödinger functional boundary conditions~\cite{Fritzsch:2013je}. 
This allows us to control the leading order discretization errors 
and hence improve the continuum limit. 
Preliminary results of this study were published in~\cite{Rantaharju:2014ila}.
Schrödinger functional boundaries also make it possible to measure 
the mass anomalous dimension 
as described in more detail in~\cite{suorsa}.

\section{Methods and Results}
We study the model using a HEX smeared~\cite{Capitani:2006ni}, 
clover improved Wilson fermion action 
and a partially smeared plaquette gauge action. 
The full action can be written as 
\begin{align*} S =
    (1-c_g)S_G(U) + c_g S_G(V) + S_F(V) + c_{SW} \delta S_{SW}(V),
\end{align*} 
where $V$ is the smeared gauge field 
and $U$ is the unsmeared one. 
As a result of the smearing, 
the action is non-perturbatively improved to order $a$ 
with the Sheikholeslami-Wohlert coefficient $c_{SW}\approx1$ 
and we simply choose $c_{SW}=1$. 
The gauge action smearing, tuned by the coefficient $c_g$, 
removes the unphysical bulk phase transition 
from the region of interest in the parameter space. 
In our case it is sufficient to choose $c_g=0.5$.

We also use Schrödinger Functional boundary conditions, 
which enables us to measure the mass anomalous dimension of 
the model from the same dataset. 
The gauge fields are set to unity 
and the fermion fields are set to zero at time slices $x_0=0,L$ 
on a lattice of size $L^4$:
\begin{alignat*}{2} 
    &U_k(x) = V_k(x) = 1, &&~~~~\psi(x) = 0  \text{~~ when ~~} x_0=0,L \\ 
	&U_\mu(x+L\hat k) = U_\mu(x), &&~~~~V_\mu(x+L\hat k) = V_\mu(x)\\ &\psi(x + L\hat k) = \psi(x). 
\end{alignat*} 
Here $k=1,2,3$ labels one of the spatial directions.

We use the gradient flow method to measure the running coupling. 
The gradient flow is defined 
by introducing a fictitious flow time $t$ 
and studying how the fields evolve according to flow equation:
\begin{align*} 
  \partial_t B_{t,\mu} &= D_{t,\mu} B_{t,\mu\nu}, \\
  B_{0,\mu} &= A_\mu\\ G_{t,\mu\nu} &= \partial_\mu B_{t,\nu} -
               \partial_\nu B_{t,\mu} + \left[ B_{t,\mu},B_{t,\nu} \right].
\end{align*} 
Here $B_{t,\mu}$ is the flow field parametrized 
by the flow time $t$, and $A_\mu$ is the original gauge field. 
In this study we evolve the flow fields using the Symanzik gauge action. 
The flow smears the gauge field over a radius $\sqrt{8t}$,
removing the UV divergences 
and automatically renormalizing gauge invariant observables~\cite{Luscher:2011bx}.

The coupling is measured from the evolution of the field strength
\begin{align*} 
    \left<E(t)\right> &= \frac 14 \left<G_{\mu\nu}(t)G_{\mu\nu}(t)\right>. 
\end{align*} 
To the leading order in perturbation theory in $\overline{\text{MS}}$ scheme, 
it has the form~\cite{Luscher:2010iy} 
$ \left<E(t)\right> = \mathcal{N} g^2/t^2 + \mathcal{O}(g^4).$ 
Since the translation symmetry is broken by the boundary conditions, 
we measure the field strength only at the middle time slice. 
The dimensionless renormalized quantity $t^2<E(t)>$ can 
then be used to define the renormalized coupling~\cite{Fodor:2012qh}:
\begin{align} \label{gf_coupling} 
    &g^2_{GF}(\mu) = \mathcal{N}^{-1}t^2 \left < E(t) \right>\vert_{x_0=L/2\,,\,t=1/8\mu^2}\,, 
\end{align} 
where the normalization factor $\mathcal{N}$ 
for these boundary conditions has been calculated in~\cite{Fritzsch:2013je}. 
To be free of both lattice artifacts and finite volume effects 
the scale $\mu$ must be within $1/L\ll\mu\ll1/a$. 
To ensure this we relate the lattice scale 
with the renormalization scale as $\mu^{-1}=cL$. 
Unless otherwise indicated we use $c=0.4$ in our analysis below.

We measure the coupling with several physical lattice sizes 
and bare couplings in order to quantify 
the running using the step scaling function~\cite{Luscher:1993gh}: 
\begin{align} \label{lat_step} 
    \Sigma(u,\A/L) &= \left . g_{GF}^2(g_0,2L/\A) \right|_{g_{GF}^2(g_0,L/\A)=u}\\\nonumber 
	\sigma(u) &= \lim_{\A\rightarrow 0} \Sigma(u,\A/L). \nonumber 
\end{align} 
The step scaling function describes how the coupling evolves 
when the linear size of the system is increased.

\begin{figure}
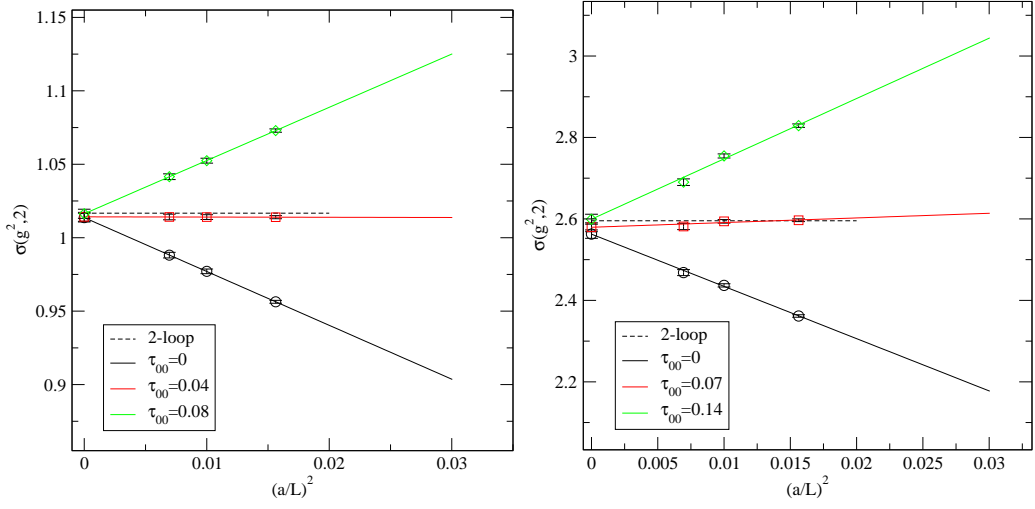
 
  \includegraphics[width=0.45\textwidth]{tautest1.eps}
  \includegraphics[width=0.45\textwidth]{tautest25.eps} 
  \caption[b]{ 
              The continuum limit~\eqref{contfit} 
			  with several values of the flow time
              correction $\tau_0$ with $u=1$ (left) and $u=2.5$ (right).
  }
  \label{fig:sigma_fit_tau} 
\end{figure} 
\begin{figure}
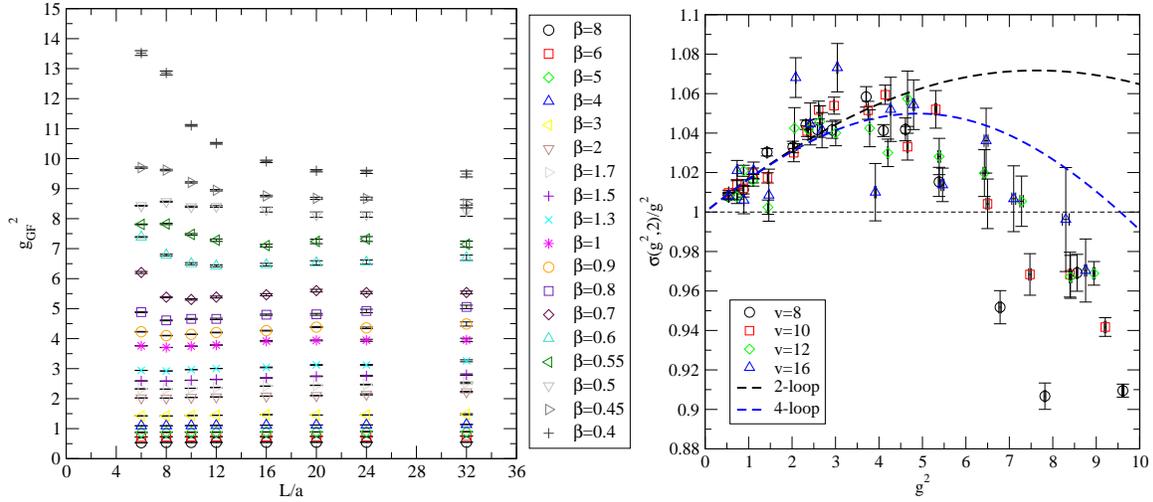

  \includegraphics[width=0.55\textwidth]{g2c04t06.eps}
  \includegraphics[width=0.45\textwidth]{latssf.eps} 
  \caption[b]{ 
              The plot on the left shows the gradient flow coupling~\eqref{gf_coupling}
              measured at each $\beta$ and $L/a$ at $c=0.4$. 
			  The plot on the right shows 
			  the lattice step scaling function~\eqref{lat_step} for these couplings. 
			  For both pictures $\tau_0$ correction~\eqref{taufunc} was used.
  }
  \label{fig:g2_lat_meas} 
\end{figure}

As was observed in~\cite{Rantaharju:2013bva}, 
the gradient flow coupling suffers from large order $(a/L)^2$ errors. 
These can be alleviated with a correction $\tau_0$ to the lattice value of 
the flow time~\cite{Cheng:2014jba} 
\begin{align} \label{taucorrection} 
    g^2_{GF} &= \mathcal{N}^{-1} t^2 \left < E(t+\tau_0 a^2) \right> 
	         = \mathcal{N}^{-1} t^2 \left < E(t) \right> 
	         + \mathcal{N}^{-1} t^2 \left < \frac{\partial E(t)  }{ \partial t} \right> {\tau_0} a^2. 
\end{align} 
The effect of the correction on the step scaling function 
can be seen in figure~\ref{fig:sigma_fit_tau}. 
As can be seen, 
$\tau_0$ does change the slope 
but not the continuum limit of the step scaling function. 
In the following results we choose a functional form 
that approximately removes the $\mathcal{O}(a^2)$ effects. 
For the $c=0.3-0.5$ we use 
\begin{align} \label{taufunc} 
    \tau_0 = 0.06 \log (1+g^2). 
\end{align} 
The measured couplings $g^2_{GF}$ 
and lattice step scaling function $\Sigma$ 
with this $\tau_0$ correction are shown in figure~\ref{fig:g2_lat_meas}.
%
The continuum limit of the step scaling function has to be evaluated at
constant coupling $g^2$. 
We use two different methods for this.
First method uses the traditional polynomial interpolation:
\begin{align} \label{betafitfun} 
    g^2_{GF}&(g_0,\A/L) = \sum_{i=1}^m  a_i g_0^{2i} \,, \;\; m=6 
\end{align}
%
\begin{figure}[t] \center
  \includegraphics[width=0.45\textwidth]{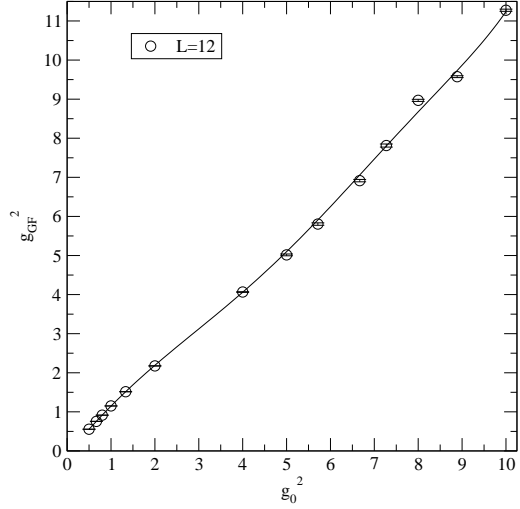} 
  \caption[b]{ 
              The gradient flow coupling and the interpolating function~\eqref{betafitfun}.
  }
  \label{fig:g2_interpolate} 
\end{figure} 
\begin{figure}[t]
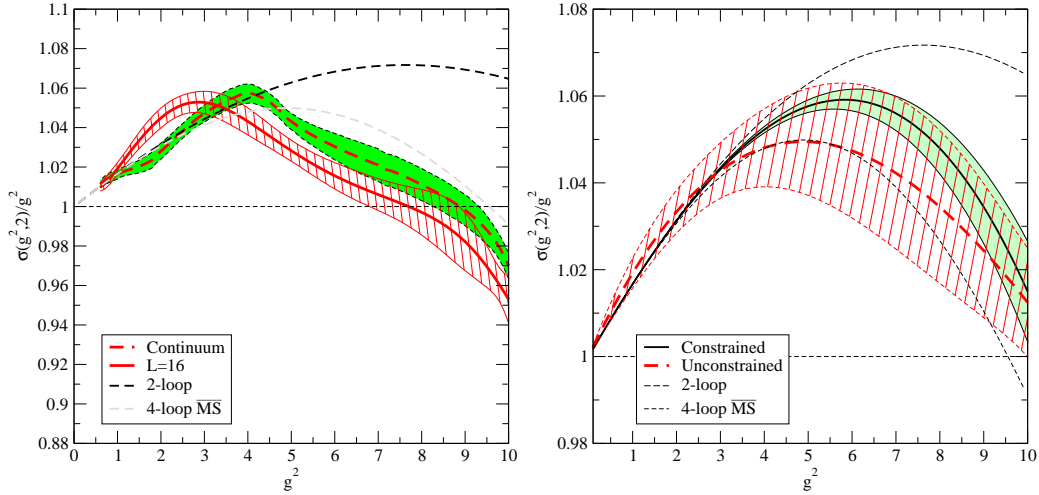

  \includegraphics[width=0.45\textwidth]{contssf.eps}
  \includegraphics[width=0.45\textwidth]{meth2.eps} 
  \caption[b]{ 
              The step scaling function with $c=0.4, \tau_0 = 0.06\log(1+u) $ 
			  calculated with two continuum extrapolation methods. 
			  On the left: traditional coupling interpolation~\eqref{betafitfun} 
			  for the biggest lattice and continuum extrapolation~\eqref{contfit}. 
			  on the right: power series~\eqref{contfit2} for $(a/L)^2+(a/L)^3$. 
			  The solid line is constrained to perturbative results, 
			  while the hashed line is not. 
			  4-loop $\overline{\text{MS}}$ results are show as a guide for the eye.
  }
  \label{fig:sigma_lat_cont} 
\end{figure}
shown in figure~\ref{fig:g2_interpolate}. 
The continuum limit $\sigma(u)$ can 
then be approximated by fitting to a function of a form:
\begin{align} \label{contfit} 
    &\Sigma(u,\A/L)  = \sigma(u) + c_2(u) (a/L)^2. 
\end{align} 
This method does not utilize the fact 
that the step scaling $\sigma$  must follow the universal perturbative 2-loop prediction.
As a second method, 
we carry out a continuum extrapolation, 
where we present $\sigma(u)$ as truncated power series, 
parametrize the discretization errors as series in $u$ 
and do a single fit to all measured data~\cite{Rantaharju:2015yva}. 
This second method has the form: 
\begin{align} \label{contfit2} 
    \Sigma(u,a/L) &= \sigma(u)+\sum_{k=2}^{n_a}f_k(u)\frac{a^k}{L^k} \\ \nonumber 
	\sigma(u)     &= 1+ \sum_{i=1}^n s_i u^i \\ \nonumber 
	f_k(u)        &= \sum_{l=0}^{n_f^k} c_{k,l}u^l\,, 
\end{align} 
where $c_i$ and $c_{k,l}$ are fit parameters, 
$s_i$ are determined from universal 2-loop $\beta$-function i.e. $s_1=2 b_0 \ln(2)$. 
We set $n=3$, $n_a=3$ and $n^2_f=n^3_f=4$ 
in order to minimize $\chi^2/\text{d.o.f}$ of the fit.
These two methods are compared in~\ref{fig:sigma_lat_cont}. 
They both follow the perturbative value closely up to $g^2=4$ 
and then diverge towards a fixed point. 
While the 4-loop $\overline{\text{MS}}$ result is scheme dependent, 
the nonperturbative results obtained 
with either of the two methods behave similar to it. 
The power series method seems to give IRFP 
at a bit higher value than the traditional approach.
\section{Conclusions} 
We have studied the running coupling in the SU(2) lattice gauge theory 
with 8 fermions in the fundamental representation.
Gradient flow algorithm with Schrödinger functional boundaries 
and flow time correction~\eqref{taucorrection} seems to work well 
even on relatively high couplings. 
The calculated step scaling function shows a clear indication of a fixed point. 
We have also measured the mass anomalous dimension 
using the mass step scaling and spectral density methods. 
The results are reported in~\cite{suorsa}.
%
%
\section{Acknowledgments}
This work is supported 
by the Academy of Finland grants 267842, 134018 and 267286, 
the Danish National Research Foundation DNRF:90 grant 
and by a Lundbeck Foundation Fellowship grant. 
T.K. is also funded by the Danish Institute for Advanced Study, 
T.R. by the Magnus Ehrnrooth foundation 
and J.M.S. by the Jenny and Antti Wihuri foundation.
The simulations were performed 
at the Finnish IT Center for Science (CSC) in Espoo, Finland, 
on the Fermi supercomputer at Cineca in Bologna, Italy 
and on the K computer at Riken AICS in Kobe, Japan. 
Parts of the simulation program have been derived 
from the MILC lattice simulation program~\cite{MILC}.

\end{document}